\documentclass[twocolumn,pra,showpacs,preprintnumbers,amsmath,amssymb,superscriptaddress]{revtex4}

\usepackage{graphicx}
\usepackage{dcolumn}
\usepackage{bm}

\begin{document}


\title{Mott Transitions of Three-Component Fermionic Atoms with Repulsive Interaction in Optical Lattices}

\author{Kensuke Inaba}%
\affiliation{NTT Basic Research Laboratories, NTT Corporation, Atsugi 243-0198, Japan}
\affiliation{JST, CREST, Chiyoda-ku, Tokyo 102-0075, Japan}

\author{Shin-ya Miyatake}
\affiliation{Department of Applied Physics, Osaka University, Suita, Osaka 565-0871, Japan}

\author{Sei-ichiro Suga}
\affiliation{Department of Materials Science and Chemistry, University of Hyogo, Himeji 671-2280, Japan}

\date{\today}

\begin{abstract}
 We investigate the Mott transitions of three-component (colors) repulsive fermionic atoms in optical lattices using the dynamical mean field theory. 
We find that for SU(3) symmetry breaking interactions the Mott transition occurs at incommensurate half filling. As a result, a characteristic Mott insulating state appears, where paired atoms with two different colors and atoms with the third color are localized at different sites. 
We also find another Mott state where atoms with two different colors are localized at different sites and atoms with the third color remain itinerant. 
We demonstrate that these exotic Mott phases can be detected by experimental double occupancy observations. 

\end{abstract}
\maketitle

Ultracold fermionic atoms loaded in optical lattices have been providing ideal stages for studying fundamental problems related to strong correlation effects \cite{Bloch,Bloch2005,Jaksch,Morsch}. 
The Mott transition, which is one of the most important phenomena as regards correlated electron systems, was observed in $^{40}{\rm K}$ fermionic atoms \cite{Jordens2008,Schneider2008}. 
The research has been extended to the topics about the phase transition of multicomponent fermionic atoms. 
Recently, the Mott transition of six-component repulsive $^{173}$Yb fermionic atoms in an optical lattice was investigated by double occupancy measurements of the site using the photo association spectroscopy \cite{Taie}. 
Interestingly, one can create the odd-number-component systems, where novel features are expected to appear.
In fact, degenerate three-component fermionic gases have been succsessfully created \cite{Ottenstein} and intensively investigated regarding the Efimov state \cite{Efimov}.

The Mott transition of three-component fermionic atoms in optical lattices was investigated for isotropic repulsive interactions with SU(3) symmetry \cite{Gorelik}.
It was shown that the Mott transition occurs at commensurate $1/3$ and $2/3$ fillings, but never occurs at incommensurate half filling. 
However, in a recent study, we showed that the Mott insulating state appear even at incommensurate half filling for SU(3) symmetry breaking interactions \cite{Miyatake1}.
We also showed that fluctuation effects induce the color selective Mott transition (CSMT) state, where atoms with two different colors are localized at different sites and the third color atoms are itinerant.
By analogy with strongly correlated electron systems \cite{Georges1996, Kotliar2004}, the Mott transition states are considered to extend at finite temperatures. 
Therefore, they are possibly observed in experiments.
However, the reason for the appearance of the Mott transitions at half filling and the characteristics of the Mott transition states have not been unveiled yet.

In this paper, we investigate the Mott transition of repulsively interacting three-component fermionic atoms in optical lattices. 
From the calculation of the double occupancy which can be observed experimentally \cite{Jordens2008,Schneider2008,Taie}, we conclude that paired atoms with two different colors and atoms with the third color are localized at different sites in the Mott insulating state. 
We call this Mott insulating state a paired Mott insulator (PMI). 
This reminds us of the Mott insulating state of Bose-Fermi mixtures \cite{BFmix}, even though we considere the repulsive interactions. From systematic calculations, rich phase diagrams of the PMI, CSMT, and Fermi liquid (FL) phases are obtained in terms of the anisotropy of the repulsive interactions. We demonstrate that these phases can be distinguished experimentally by observing double occupancy. 
We show that the phase transitions from the CSMT state to the PMI state are effectively described by the Falicov-Kimball model, which has been studied as a minimal model for the metal-insulator transition in $d-$ and $f-$electron systems \cite{FK}. 
We thus expect that the three-component optical lattice system allows us to simulate widespread phenomena related to the Bose-Fermi mixtures and the correlated electrons with some orbitals.

In accordance with the conventional model for cold atoms in optical lattices \cite{Jaksch}, we set the nearest-neighbor hopping and the on-site interaction between atoms with different colors. The low-energy properties can be described by the following Hamiltonian, 
\begin{eqnarray}
{\cal H}&=-&\sum_{\langle i,j \rangle}\sum_{\alpha=1}^{3}
       \left( t+\mu_\alpha \delta_{i,j} \right)a^\dag_{i\alpha} a_{j\alpha} 
   + \frac{1}{2}\sum_{i}\sum_{\alpha\not=\beta} 
       U_{\alpha\beta} n_{i \alpha} n_{i \beta},   \nonumber \\
\label{eq_model}
\end{eqnarray}
where the subscript $\langle i,j \rangle$ is the summation over the nearest neighbor sites, $a^\dag_{i\alpha} (a_{i\alpha})$ is the fermionic creation (annihilation) operator for the state with color $\alpha$ in the $i$th site and $n_{i\alpha}=a^\dag_{i\alpha}a_{i\alpha}$. 
Here, $t$ denotes the hopping integral, 
$\mu_\alpha$ is the chemical potential for the atom with color $\alpha$, and 
$U_{\alpha\beta}$ is the repulsive interaction between two atoms with colors $\alpha$ and $\beta$. 
The hopping integral $t$ is used in units of energy. 
The chemical potential is set at $\mu_{\alpha}=(U_{\alpha \beta}+U_{\gamma \alpha})/2$ so that particle-hole symmetry can be satisfied for each color and the atoms with each color achieve a balanced population at half filling. 
We set $U_{12}=U$, $U_{23}=U'$, and $U_{31}=U''$. 
We assume a homogeneous optical lattice and neglect the confinement potential for a first approximation.

To obtain a self-consistent solution in the DMFT calculations, we employ the two-site DMFT method as a solver \cite{Potthoff2001}. 
We use a semicircular density of states obtained for the infinite-dimensional bipartite Bethe lattice.
We calculate the quasiparticle weight $Z_{\alpha}$. 
To investigate the characteristics of a Mott transition, we further calculate the occupation rate of atoms with colors $\alpha$ and $\beta$ in the same site, 
$D_{\alpha \beta} = \langle n_\alpha n_\beta \rangle$.

\begin{figure}[htb]
\begin{center}
\includegraphics[scale=0.3]{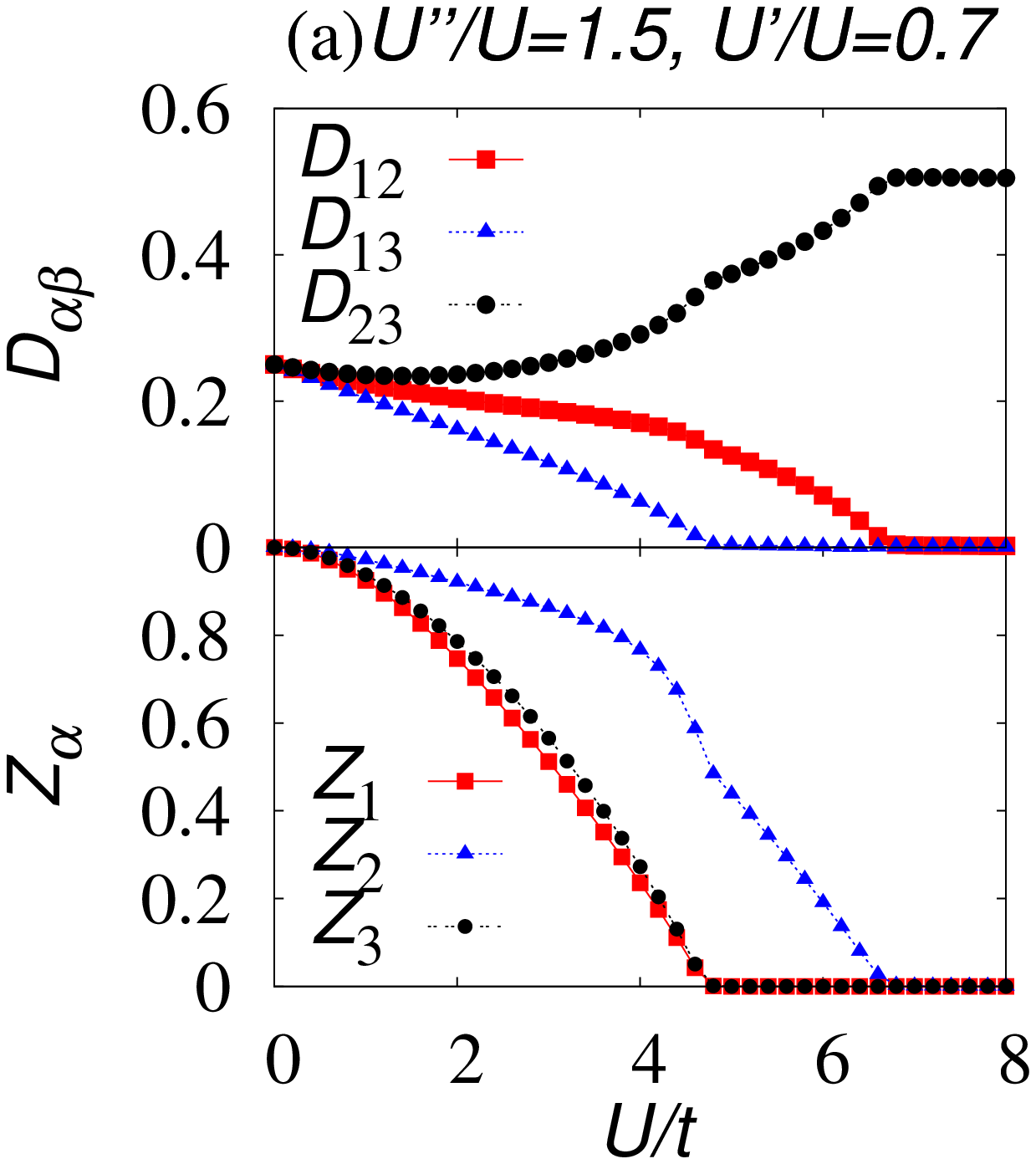}
\includegraphics[scale=0.3]{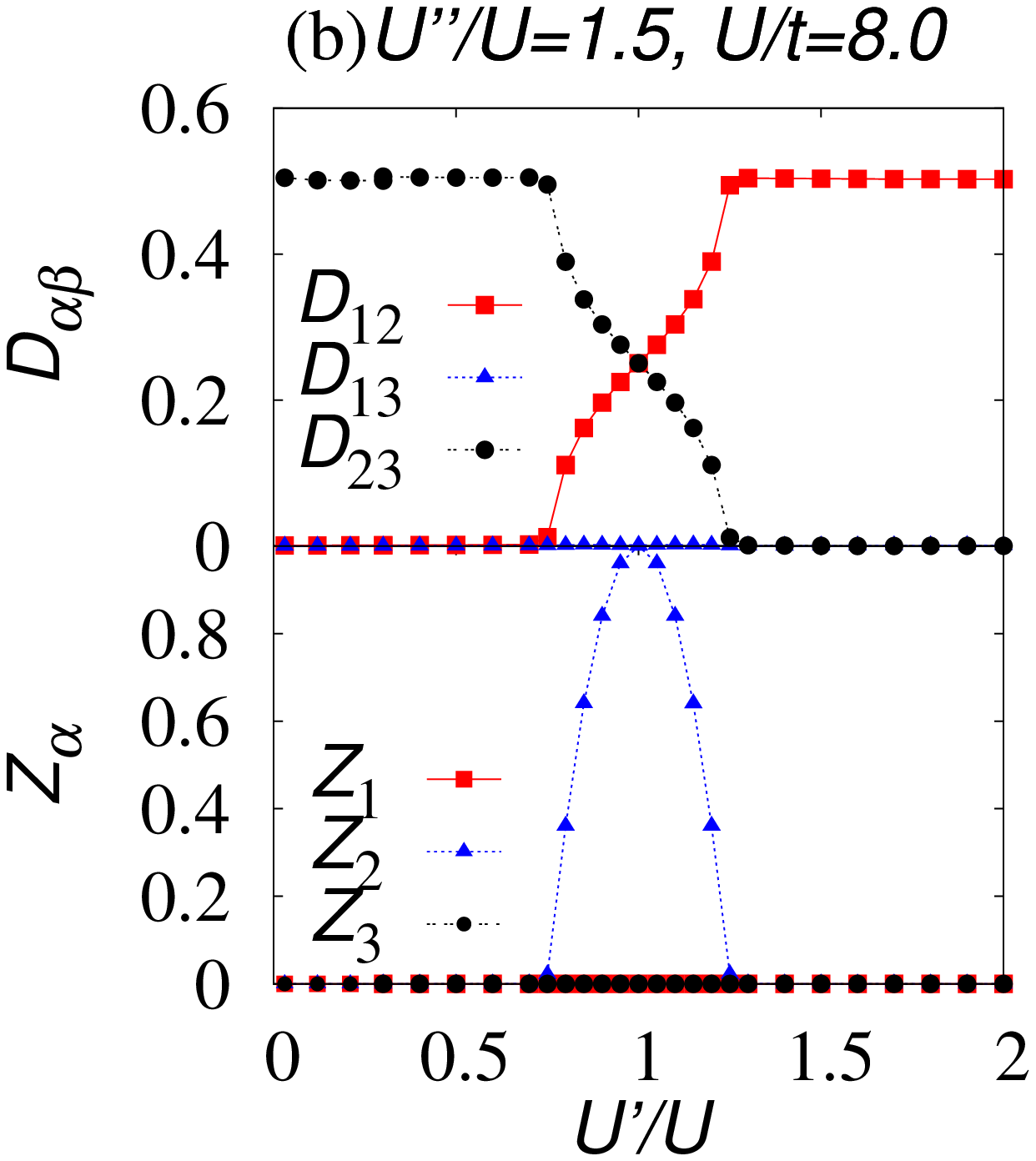}
\caption{(Color online)  Quasiparticle weight $Z_{\alpha}$ and double occupancy $D_{\alpha \beta}$ (a) for $U''/U = 1.5$ and $U'/U = 0.7$ as a function of $U/t$, and (b) for $U''/U = 1.5$ and $U/t = 8.0$ as a function of $U'/U$. 
}
\label{fig4}
\end{center}
\end{figure}

In Fig. \ref{fig4}(a), we show the results for $U''/U = 1.5$ and $U'/U = 0.7$ as a function of $U/t$. 
As $U/t$ is increased, $Z_{\alpha}$ decreases monotonically. 
 $Z_1$ and $Z_3$ become zero at $U_{c1}/t=4.7$, while $Z_2$ becomes zero at $U_{c2}/t=6.7$. 
$D_{12}$ and $D_{13}$ decrease and become zero at $U_{c1}$ and $U_{c2}$, respectively. On the other hand, $D_{23}$ increases towards $1/2$ at $U_{c2}$ with increasing $U/t$. 
From the results for $Z_{\alpha}$, we find that the FL state appears for $U<U_{c1}$. 
For $U_{c1}<U<U_{c2}$, atoms with colors 1 and 3 are localized at different sites as a result of the Mott transition caused by the strongest $U''$. By contrast, atoms with color 2 are itinerant throughout the system as renormalized fermions. 
This is the CSMT state. 
In the following, we call the CSMT state with itinerant color-$\alpha$ atoms CSMT-$\alpha$. 
In CSMT-$\alpha$, atoms with colors $\beta$ and $\gamma$ are localized at different sites owing to the Mott transition, leading to $D_{\beta \gamma}=0$. Other $D_{\gamma \alpha}$ and $D_{\alpha \beta}$ take nonzero values depending on the interactions. 
For $U>U_{c2}$, the Mott insulating state appears. 
In this state, in order to avoid the collision induced by two stronger interactions, color-2 and 3 atoms are localized at the same sites, while color-1 atoms are only localized at different sites. 
We call this Mott insulating state PMI-$23$. 
We can understand this feature from the behavior of the double occupancy $D_{\alpha \beta}$. 
In the Mott phases, $D_{\alpha \beta}$ can be rewritten as $D_{\alpha \beta}= \sum_\phi \rho_\phi d^\phi_{\alpha\beta}$, 
where $\rho_\phi$ is the weight of the local eigenstate $|\phi\rangle$ and $d^\phi_{\alpha\beta} =\langle \phi|n_\alpha n_\beta|\phi\rangle$. 
When the paired localized state of color-$\alpha$ and $\beta$ atoms $|\phi_{\alpha \beta}\rangle$ or the single localized state of a color-$\gamma$ atom $|\phi_{\gamma}\rangle$ appears randomly in each site,  we obtain $(d^\phi_{\alpha \beta},d^\phi_{\beta \gamma},d^\phi_{\gamma \alpha})=(1,0,0)$ and $(0,0,0)$, respectively.
These two localized states are distributed with an equal probability, $\rho_{\phi_{\alpha \beta}}=\rho_{\phi_{\gamma}}=1/2$. 
We thus obtain $(D_{\alpha \beta},D_{\beta \gamma},D_{\gamma \alpha})=(1/2,0,0)$ as shown in Fig.\,\ref{fig4}\,(a).

The CSMT state and the PMI are characteristic Mott transition states of three-component repulsive fermionic atoms in optical lattices. 
In the CSMT state, the itinerant atoms are influenced by the potential scattering of the randomly distributed localized atoms with two different colors, which drives the itinerant atoms to the PMI. 
This driving mechanism is different from that of the orbital-selective Mott transition \cite{Koga}, which appears in strongly correlated electron systems described by the two-band Hubbard model with different bandwidths.

In Fig. \ref{fig4}(b), we show $Z_{\alpha}$ and $D_{\alpha \beta}$ as a function of $U'/U$ for $U''/U=1.5$ and $U/t=8.0$. 
In these parameters, $Z_1$, $Z_3$, and $D_{13}$ are equal to zero for $0 \leq U'/U \leq 2$, indicating that atoms with colors 1 and 3 obey the Mott transition driven by the strongest repulsion $U''$. 
For $U'/U<0.7$ and $U'/U>1.3$, we find $D_{12}=0$ and $D_{23}=0$, respectively, in addition to the above results. The results indicate that for $U'/U<0.7$ ($U'/U>1.3$) color-2 and 3 (color-1 and 2) atoms with the weakest $U'$ ($U$) are localized at the same sites and form pairs, yielding PMI-23 (PMI-12).  
For $0.7<U'/U<1.3$, color-2 atoms are itinerant owing to frustration effects caused by the nearly balanced $U$ and $U'$. Accordingly, CSMT-2 appears. 
At $U'/U=1$, $Z_2$ becomes unity, which indicates that the renormalization effects vanish and atoms with color 2 behave as free fermions.

\begin{figure}[htb]
\begin{center}
\includegraphics[scale=0.3]{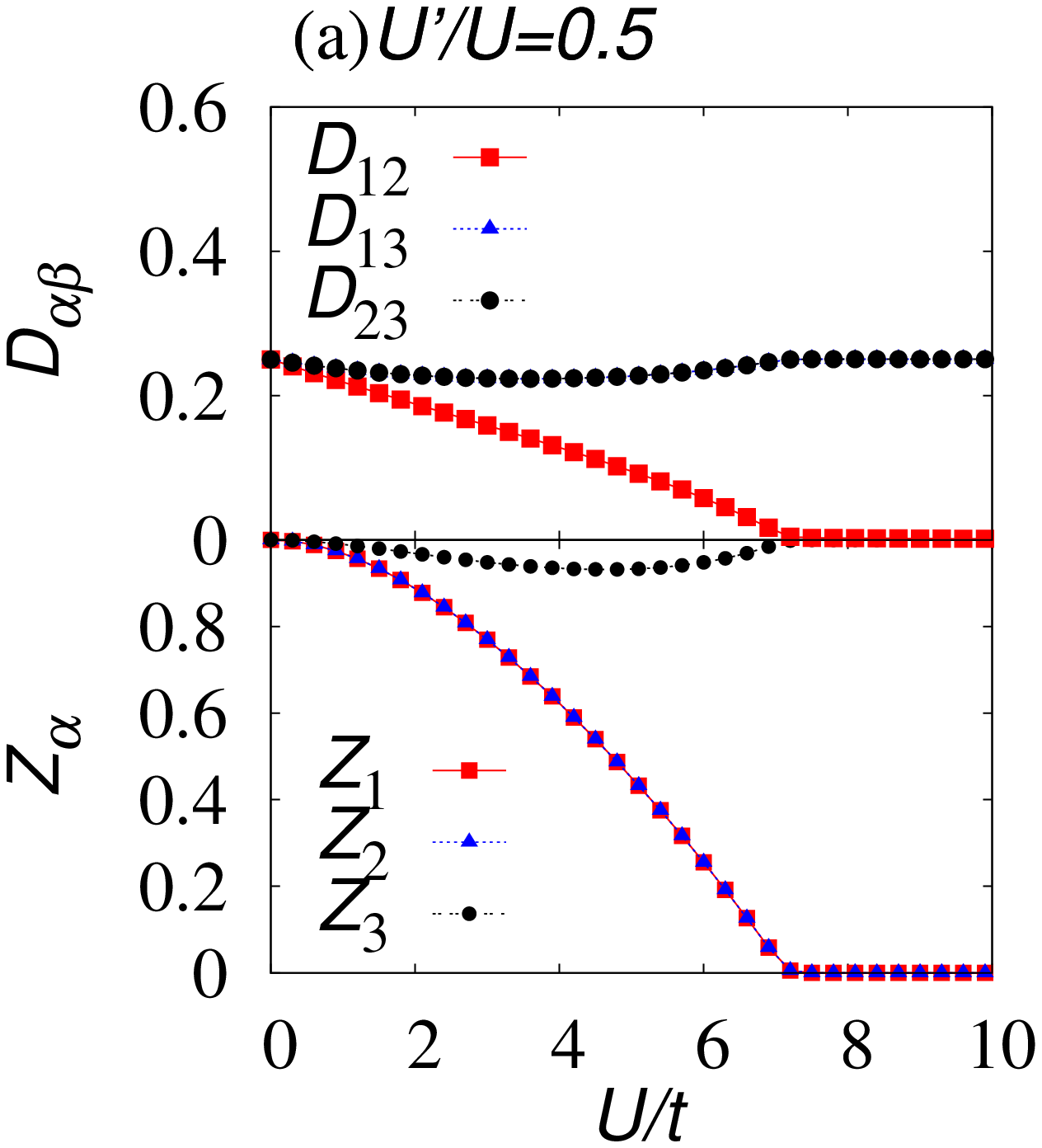}
\includegraphics[scale=0.3]{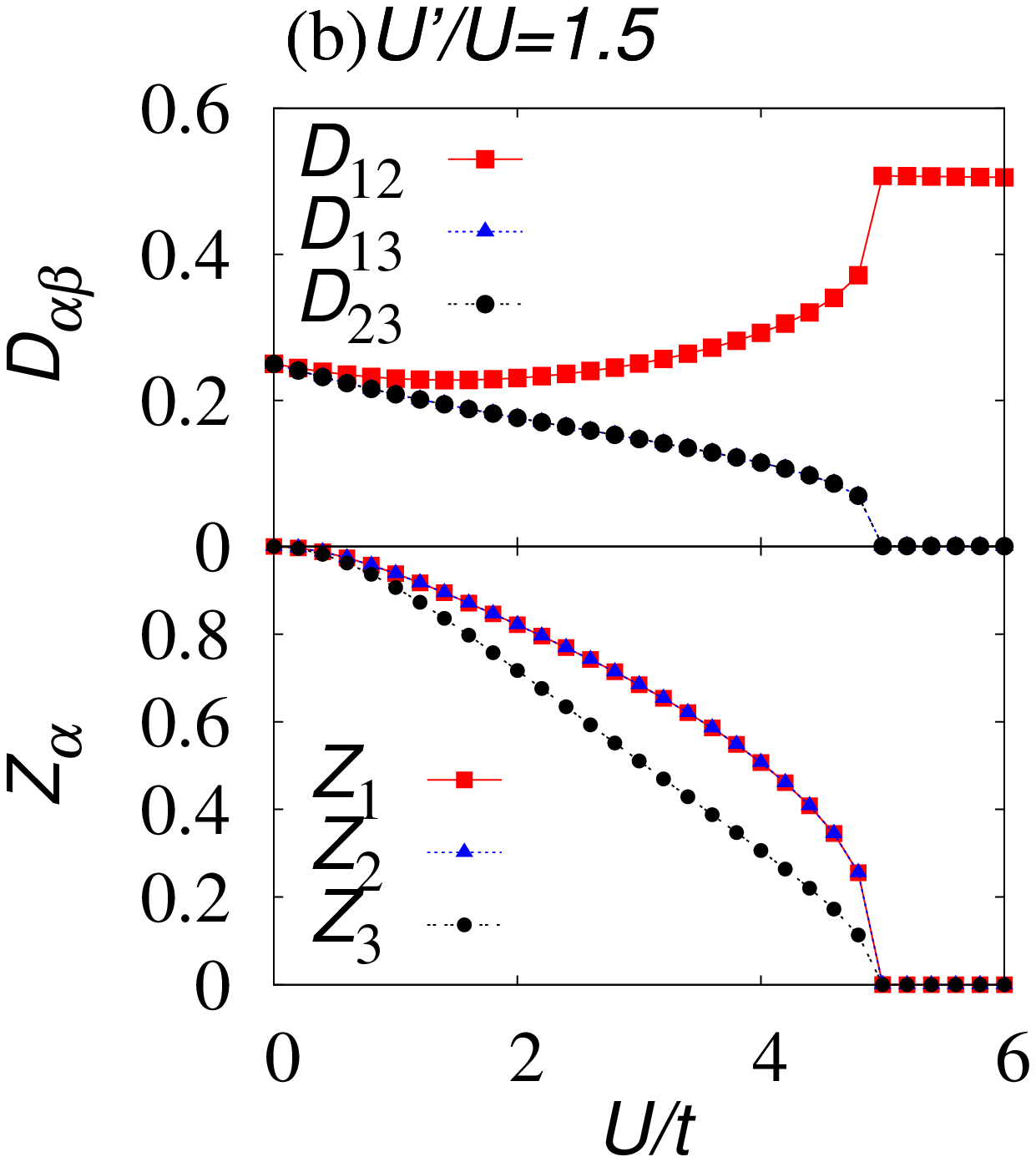}
\caption{ (Color online) In  $U \neq U'=U''$, quasiparticle weight $Z_{\alpha}$ and double occupancy $D_{\alpha \beta}$ for (a) $U'/U = 0.5$ and (b) $U'/U = 1.5$ as a function of $U/t$. Since $U'=U''$, we obtain $Z_1=Z_2$ and $D_{13}=D_{23}$.}
\label{fig2}
\end{center}
\end{figure}
To clarify the strong fluctuation effects induced by the balanced interactions, we investigate the characteristics when two of the three repulsive interactions are the same. 
For this purpose, we set $U \neq U' = U''$. 
 In Fig. \ref{fig2}(a), we show $Z_{\alpha}$ and $D_{\alpha \beta}$ for $U'/U=0.5$. 
As $U/t$ increases, $Z_1$ and $Z_2$ decrease monotonically and become zero at $U_c/t=7.1$. This Mott transition for color-1 and 2 atoms is caused by the strongest $U$. 
On the other hand, $Z_3$ first decreases and then increases nontrivially towards unity at the same $U_c/t$. Corresponding to these renormalization effects for $Z_3$, $D_{23}$ and $D_{13}$ show nonmonotonic $U/t$ dependence and become $1/4$ in $U \geq U_c$. 
The results indicate that there is a quantum phase transition from the FL to CSMT-3. 
Since $U'=U''$, two paired localized states $|\phi_{1,3}\rangle$ and $|\phi_{2,3}\rangle$, and two solely localized states $|\phi_{2}\rangle$ and $|\phi_{3}\rangle$ emerge with the same probability in each site in CSMT-3. 
We evaluate $d_{\alpha \beta}^{\phi}$ and $\rho_{\phi}$ in the same way as mentioned above, resulting in $(D_{12},D_{13},D_{23})=(0,1/4,1/4)$. 
This CSMT-3 never obeys the transition to the PMI and the itinerant color-3 atoms behave as free fermions because of the strong frustration effects induced by the balanced $U'=U''$. 
The nontrivial increase of $Z_3$ in the vicinity of a transition point is due to these frustration effects.

For $U'/U=1.5$, three quasiparticle weights decrease monotonically and become zero at $U_c/t=5.0$ as shown in Fig. 2(b). 
Since $D_{13}$, $D_{23}$, and $Z_\alpha$ vanish and $D_{12}$ becomes 1/2, 
we find a direct quantum phase transition from the FL to PMI-12 at $U_c/t=5.0$.

\begin{figure}[ht]
\begin{center}
\includegraphics[scale=0.48]{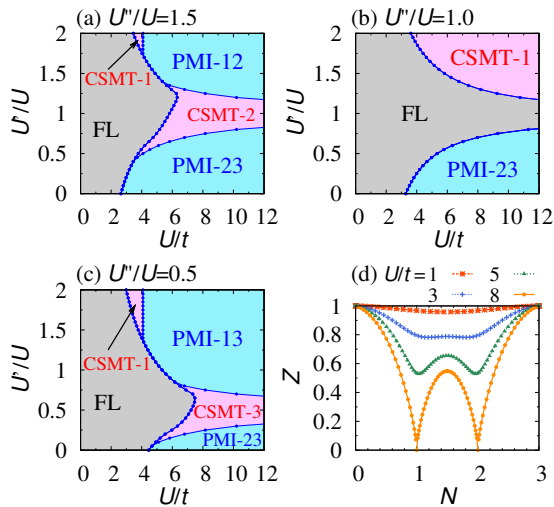}
\caption{(Color online)  Phase diagrams of the three-component repulsive fermionic atoms in optical lattices at half filling for (a) $U''/U=1.5$, (b) $U''/U=1.0$, (c) $U''/U=0.5$. 
(d): Quasiparticle weight for $U = U' = U''$ as a function of the atom number $N$ per site.  Because of the isotropic interaction, $Z \equiv Z_1 = Z_2 = Z_3$. 
}
\label{fig5}
\end{center}
\end{figure}

We perform the same calculations by changing $U/t$ and $U'/U$ systematically for a given $U''/U$. The results are summarized in the phase diagrams shown in Fig. \ref{fig5}(a), (b), and (c). The FL states appear in the small $U/t$ regions. As $U/t$ increases, there are various quantum phase transitions depending on the anisotropy of the repulsive interactions. 
For $U''/U=1.5$ and $0.5$, we find the same topological features as found in the phase diagrams. 
For the large $U/t$ regions in $U'' \neq U' \neq U$, the two different color atoms possessing the strongest interaction are localized at different sites and the third color atoms form pairs with atoms corresponding to the weakest interaction, which yields the PMI as shown in Fig. \ref{fig5}(a) and (c). 
When the two weaker interactions are nearly the same, the CSMT state appears where atoms associated with the two weaker interactions are itinerant as renormalized fermions. 
When the two stronger interactions are nearly the same, there is a direct transition from the FL to the PMI with increasing $U/t$. In the FL state close to this direct transition line, we can expect pair fluctuations to be enhanced, which may yield superconducting fluctuations under appropriate conditions \cite{M2X}. 

When the two weaker (stronger) interactions are the same, there is a transition from the FL to the PMI (CSMT state) with increasing $U/t$. 
In this case, the itinerant atoms in the CSMT state acts as free fermions. 
When the interactions are isotropic, there is no Mott transition \cite{M2X,Gorelik} and the FL remains with increasing $U/t$ as shown in Fig. \ref{fig5}(b). 
This result can also be confirmed by the quasiparticle weight as a function of the atom number $N$ per site. As shown in Fig. \ref{fig5}(d), $Z(\equiv Z_1=Z_2=Z_3)$ decreases with increasing $U/t$ and becomes zero only at $N=1$ and $2$, which correspond to the commensurate $1/3$ and $2/3$ fillings. At $N=3/2$, which corresponds to incommensurate half filling, $Z$ remains nonzero value even at $U/t=8$.

\begin{figure}[bht]
\begin{center}
\includegraphics[scale=0.31]{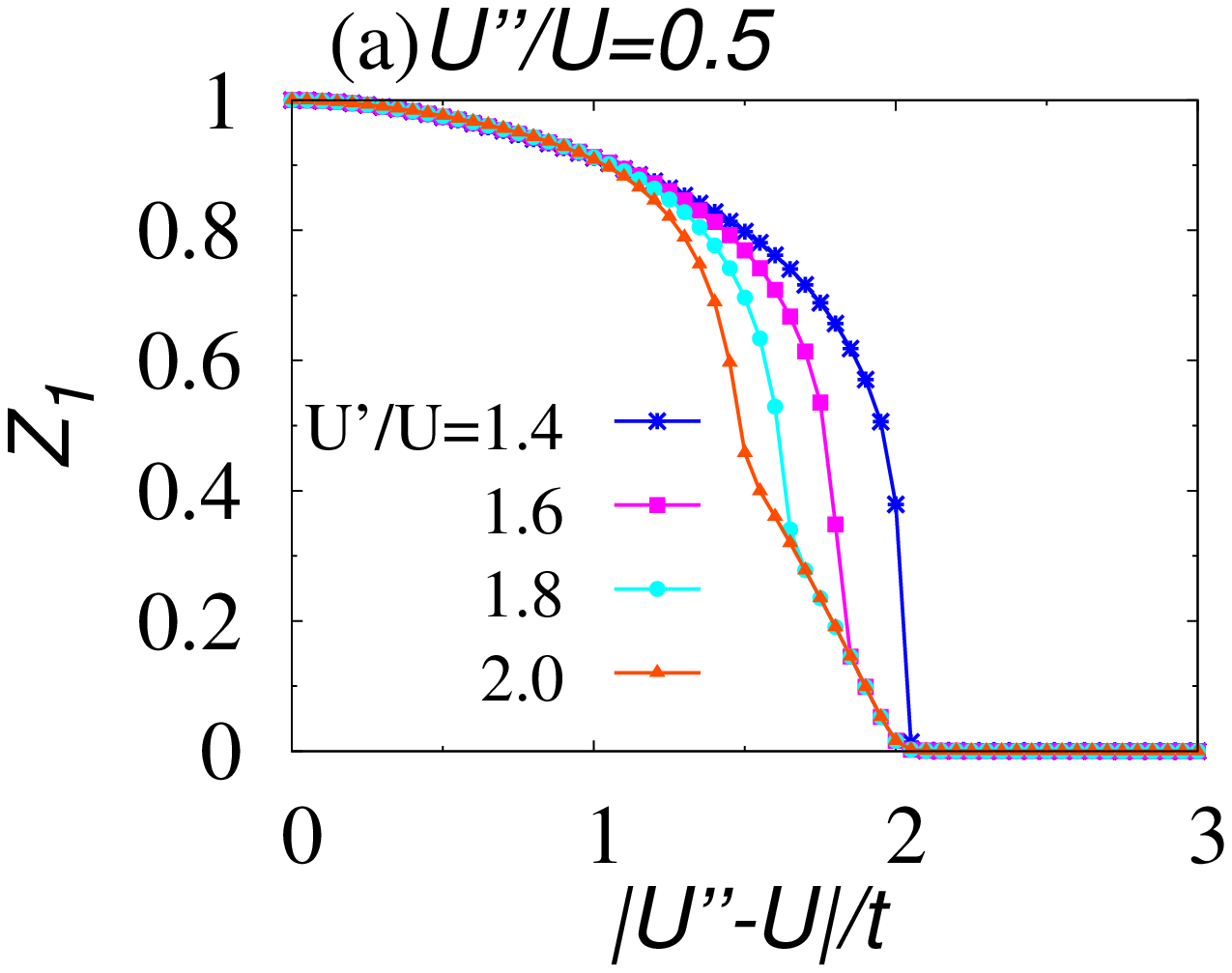}
\includegraphics[scale=0.31]{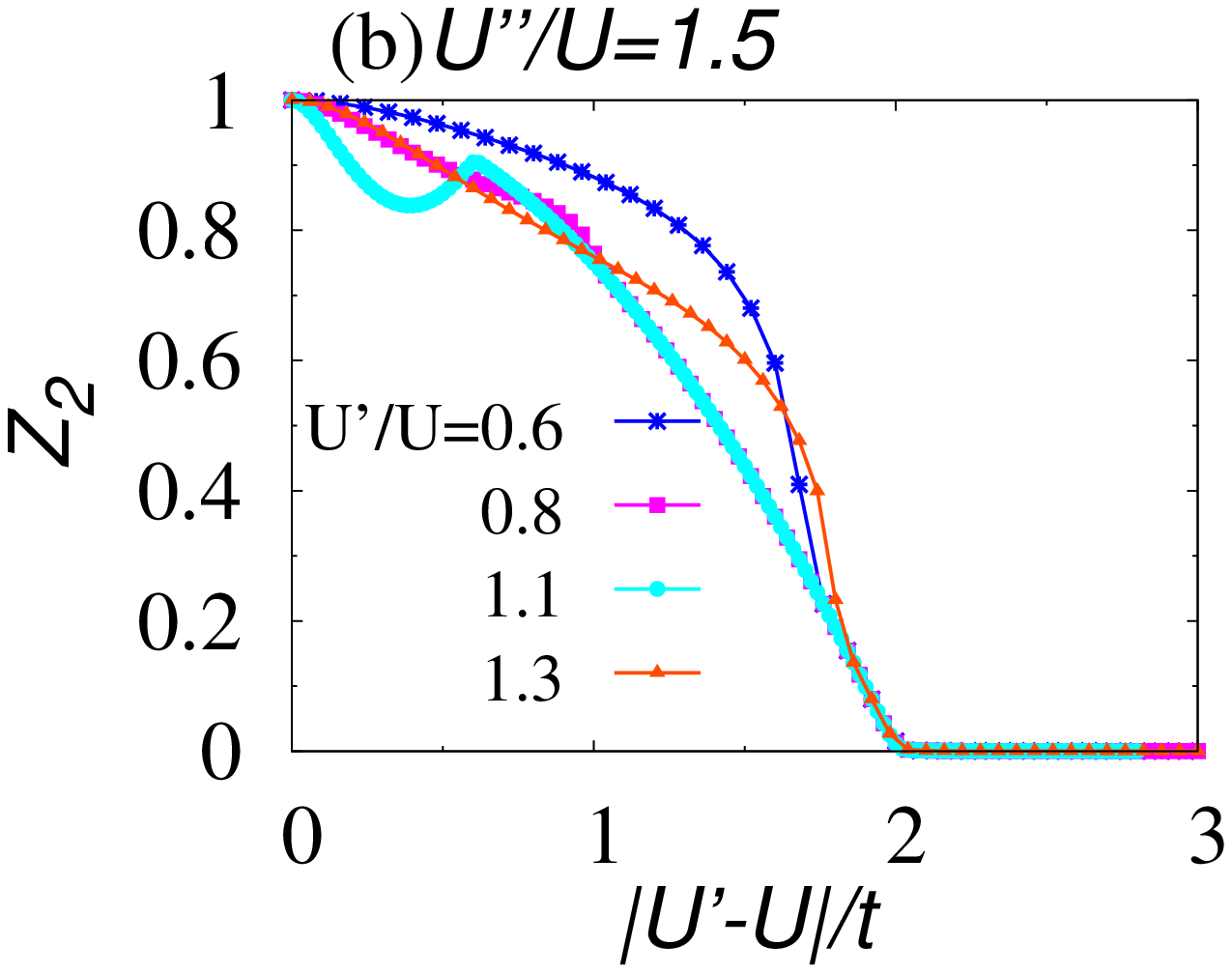}
\caption{(Color online) Quasiparticle weights (a) $Z_1$ for $U''/U=0.5$ as a function of $\tilde{U}/t=|U''-U|/t$ for $U'/U=1.4, 1.6, 1.8$, and $2.0$, and (b) $Z_2$ for $U''/U=1.5$ as a function of $\tilde{U}/t=|U'-U|/t$ for $U'/U=0.6, 0.8, 1.1$, and $1.3$. 
}
\label{fig6}
\end{center}
\end{figure}

In Fig. \ref{fig5} (c), CSMT-1 appears between the FL and PMI-13. The quantum phase transitions from CSMT-1 to PMI-13 occur at the same $U/t=4.0$ irrespective of $U'/U$. We discuss this feature in terms of an effective model. 
When two of the three color atoms are localized by the Mott transition at half filling, the effective Hamiltonian can be derived \cite{Miyatake1}, which is known as the Falicov-Kimball model. 
In deriving the effective Hamiltonian, we transform the degrees of freedom related to which color atom is localized at the site into the charge degrees of freedom of the localized spinless fermion. 
We consider the CSMT state in $U'/U \geq 1.4$ for $U''/U=0.5$. In this case, $U'$ is the strongest and color-1 atoms are itinerant. The effective Hamiltonian takes the form 
$\tilde{\cal H} = - t\sum_{\langle i,j \rangle}c^\dag_{i}c_{j} 
   - \frac{1}{2}\tilde{U} \sum_{i}\left(c^\dag_{i}c_{i}+f^\dag_{i}f_{i} \right)
   + \tilde{U}\sum_{i} c^\dag_{i}c_{i}f^\dag_{i}f_{i},  
$
where $c_{i}$ ($f_{i}$) is the annihilation operator of the spinless (localized) fermion, and $\tilde{U}=|U''-U|$ is the effective interaction between localized and spinless fermions. 
On the basis of the effective Hamiltonian, we depict $Z_1$ as a function of the effective interaction $\tilde{U}/t$ for $U'/U=1.4, 1.6, 1.8$, and $2.0$. The results are shown in Fig. \ref{fig6}(a). 
The $Z_1$ curves decrease with increasing $\tilde{U}/t$ and become zero at $\tilde{U}_c/t \sim 2.0$ irrespective of $U'/U$. 
As $\tilde{U}/t$ approaches this critical value, the $Z_1$ curves for $U'/U=1.8$ and $1.6$ join with the $Z_1$ curve for $U'/U=2.0$ at $\tilde{U}_j/t \sim 1.65$ and $1.80$, respectively. 
Because $U''=0.5U$, the $U_j/t$ and $U_c/t$ corresponding to these values can be evaluated as $U_j/t \sim 3.3$ and $3.6$, respectively, and $U_c/t \sim 4.0$. 
We compare these values with the CSMT-1 region shown in Fig. \ref{fig5}(c). We find that the boundaries between CSMT-1 and the other two phases agree well with $U_j/t$ and $U_c/t$.

Also for $U''/U=1.5$ shown in Fig. \ref{fig5}(a), we depict the quasiparticle weights as a function of the effective interaction $\tilde{U}$ for $U'/U=0.6$, $0.8$, $1.1$, and $1.3$. In this case $U''$ is the strongest, yielding $\tilde{U}=|U'-U|$ and CSMT-2. The results are shown in Fig. \ref{fig6}(b). 
As $\tilde{U}/t$ is increased, the $Z_2$ curves for $U'/U=0.6$, $0.8$, and $1.3$ join with the $Z_2$ curve for $U'/U=1.1$, and they become zero at $\tilde{U}/t=2.0$. 
We evaluate $U_j/t$ and $U_c/t$ in the same way and find that the regions $U_j/t \leq U/t \leq U_c/t$ agree well with the CSMT-2 region shown in Fig. \ref{fig5}(a). 
For $U'/U=1.1$, where two weak interactions are nearly balanced, $Z_2$ takes a nontrivial local maximum at $\tilde{U}/t \sim 0.6$. At this value, $Z_1$ and $Z_3$ become zero and there is a phase transition from the FL to CSMT-2. The nontrivial increase of $Z_2$ towards this transition indicates that the renormalization effects are suppressed by the frustration effects caused by the nearly balanced $U'$ and $U$.

We have shown that the quasiparticle weights of itinerant atoms in the CSMT state are scaled to the unified curve in terms of the Falicov-Kimball model. The obtained Mott transition point is $\tilde{U}_c/t=2.0$, which agrees with that of the Falicov-Kimball model \cite{vanDongen}. 
Judging from these findings, we conclude that the effective model well describes the CSMT state not only qualitatively but also quantitatively. 
Therefore, the present system is a tangible candidate for simulating interesting features of the Falicov-Kimball model. In addition to the Mott transition, itinerant ferromagnetism may appear below half filling \cite{FK}.

In summary, we have investigated three-component repulsive fermionic atoms in optical lattices. Depending on the anisotropy of the interactions, the FL, the CSMT state, and the PMI appear at half filling. 
We have shown that the double occupancies take characteristic values in these three states. 
A consideration of the confinement potential indicates that the domains of these three states appear as well as other state such as a band insulating state and commensurate $1/3$- and $2/3$-filling Mott insulating states. 
The characteristics of the domains can be observed via the double occupancy: For example, when two of the three double occupancies are close to zero, the PMI domain is realized. When one of the three double occupancies is close to zero, the CSMT domain is realized. 
These domains of the Mott states appear at high temperatures. At sufficiently low temperatures, other interesting domains such as the color-selective antiferromagnetic state and the color density-wave state are expected to emerge \cite{Miyatake1}. 
We hope that our results contribute to the study of the exotic Mott states of multicomponent cold fermionic atoms in optical lattices.

We thank Y. Takahashi and M. Yamashita for useful comments and valuable discussions. 
Some of the numerical computations were undertaken at the Supercomputer Center at ISSP, University of Tokyo. 
This work was supported by Grants-in-Aid for Scientific Research (C)
(No. 20540390) from JSPS and on Innovative Areas (No. 21104514) from MEXT.


\begin{thebibliography}{00}
%

\bibitem{Bloch} 
I. Bloch and M. Greiner: in {\it Advances in Atomic, Molecular, and Optical Physics}, ed. P. Berman and C. Lin (Academic Press, New York, 2005) Vol. 52, p. 1.

\bibitem{Bloch2005} 
I. Bloch,  Nat. Phys. {\bf 1}, 23 (2005).

\bibitem{Jaksch} 
D. Jaksch and P. Zoller,  Ann. Phys. (N.Y.) {\bf 315}, 52 (2005).

\bibitem{Morsch} 
O. Morsch and M. Oberthaler,  Rev. Mod. Phys. {\bf 78}, 179 (2005).

\bibitem{Jordens2008} 
R. J\"{o}rdens, N. Strohmaier, K. G\"{u}nter, H. Moritz,  and T. Esslinger,  Nature {\bf 455}, 204 (2008).

\bibitem{Schneider2008} 
U. Schneider, L. Hackerm\"{u}ller, S. Will, Th. Best, I. Blich, T. A. Costi, R. W. Helmes, D. Rasch, and A. Rosch,  Science {\bf 322}, 1520 (2008).


\bibitem{Taie}  
S. Taie, S. Sugawa, R. Yamazaki, and Y. Takahashi, in Talk of International Symposium on Physics of New Quantum Phases in Superclean Materials, Yokohama, March 2010. 


\bibitem{Ottenstein} 
T. B. Ottenstein, T. Lompe, M. Kohnen, A. N. Wenz, and S. Jochim, Phys. Rev. Lett. {\bf 101}, 203202 (2008).

\bibitem{Efimov} 
For a review, see V. Efimov, Nat. Phys. {\bf 5}, 533 (2009).




\bibitem{Gorelik}
E. V. Gorelik and N. Bl\"{u}mer, Phys. Rev A {\bf 80}, 051602(R) (2009).

\bibitem{Miyatake1} 
S. Miyatake, K. Inaba, and S. Suga, Phys. Rev. A {\bf 81}, 021603(R) (2010). 

\bibitem{Georges1996} 
A. Georges, G. Kotliar, W. Krauth, and M. J. Rozenberg, Rev. Mod. Phys. {\bf 68}, 13 (1996).

\bibitem{Kotliar2004}
G. Kotliar and D. Vollhardt, Physics Today {\bf March 2004}, 53 (2004).

\bibitem{BFmix} 
I. Titvinidze, M. Snoek, and W. Hofstetter, Phys. Rev. Lett. {\bf 100}, 100401 (2008). 

\bibitem{FK}  
L. M. Falicov and J. C. Kimball, Phys. Rev. Lett. {\bf 22}, 997 (1969). 


\bibitem{Potthoff2001}
M. Potthoff, Phys. Rev. B {\bf 64}, 165114 (2001). 

\bibitem{Koga}  
A. Koga, N. Kawakami, T. M. Rice, and M. Sigrist, Phys. Rev. Lett. {\bf
92}, 216402 (2004). 

\bibitem{M2X} 
S. Miyatake, K. Inaba, and S. Suga, Physica C (2010) in press. 

\bibitem{vanDongen}  
P. G. J. van Dongen, Phys. Rev. B {\bf 45}, 2267 (1992). 



\end{thebibliography}


\end{document}